\documentclass[12pt,preprint]{aastex}
\usepackage{natbib}

\newcommand{\msun}{M$_{\sun}$\,}
\newcommand{\myr}{M$_\odot$~yr$^{-1}$} 

\newcommand{\ha}{H$\alpha$}
\newcommand{\hb}{H$\beta$}
\newcommand{\nii}{[N{\sc ii}]}
\newcommand{\oiii}{[O{\sc iii}]}
\newcommand{\hii}{H{\sc ii}}
\newcommand{\kms}{km\,s$^{-1}$}

\shorttitle{Weak AGN in High-z Galaxies}

\shortauthors{Wright et al.~2010}

\begin{document}

\title{The Presence of Weak Active Galactic Nuclei in High Redshift Star Forming Galaxies}

\author{Shelley A. Wright\altaffilmark{1,3}, James E. Larkin\altaffilmark{2},  James R. Graham\altaffilmark{1} \& Chung-Pei Ma\altaffilmark{1} }

\begin{abstract}

We present [O{\sc iii 5007\AA}] observations of the star forming galaxy HDF-BMZ1299 (z=1.598) using Keck Observatory's Adaptive Optics system with the near-infrared integral field spectrograph OSIRIS. Using previous \ha~and \nii~measurements of the same source, we are able for the first time to use spatially resolved observations to place a high-redshift galaxy's substructure on a traditional \hii~diagnostic diagram. We find that HDF-BMZ1299's spatially concentrated nebular ratios in the central $\sim$1.5 kiloparsec (0$\farcs$2) are best explained by the presence of an AGN: log(\nii/\ha)=-0.22$\pm$0.05 and 2$\sigma$ limit of log(\oiii/\hb)$\gtrsim$0.26. The dominant energy source of this galaxy is star formation, and integrating a single aperture across the galaxy yields nebular ratios that are composite spectra from both AGN and \hii~regions. The presence of an embedded AGN in HDF-BMZ1299 may suggest a potential contamination in a fraction of other high-redshift star forming galaxies, and we suggest that this may be a source of the ``elevated" nebular ratios previously seen in seeing-limited metallicity studies.  HDF-BMZ1299's estimated AGN luminosity is L$_{H\alpha}$ = 3.7x10$^{41}$ erg s$^{-1}$ and L$_{[OIII]}$ = 5.8x10$^{41}$  erg s$^{-1}$, making it one of the lowest luminosity AGN discovered at this early epoch.  

\end{abstract}


\altaffiltext{1}{Department of Physics and Astronomy,
University of California, Berkeley, CA, 94720}
\altaffiltext{2}{Department of Physics and Astronomy,
University of California, Los Angeles, CA 90095}
\altaffiltext{3}{Hubble Postdoctoral Fellow; saw@astro.berkeley.edu}

\section{Introduction}\label{intro}

Determining metallicities for individual galaxies, galactic substructures, and numerous galaxies over a range of redshifts provides crucial constraints on our understanding of galaxy formation and evolution.  The relationship between metallicity of a galaxy and its total stellar mass is well-known in the local universe \citep{leq79}.  Recently, this empirical mass-metallicity (MMZ) relationship has been greatly enhanced with the large local (z $\lesssim$0.1) galaxy sample from the Sloan Digital Sky Survey (SDSS; \citealt{trem04}), which shows that galaxies with higher stellar masses have higher metallicities. Extending the local MMZ relationship to higher redshift has been performed by a few observations (e.g., \citealt{sav05, erb06a, liu08, maio08}) and shows some evolution between these regimes; namely, galaxies of a given stellar mass tend to be more metal poor at high redshifts than in the local universe. There have only been a handful of high redshift galaxies used for these studies, and better constraining the overall shape of the MMZ relationship and the evolutionary trend at any given cosmic epoch is imperative for our understanding of metal enrichment and galaxy formation.

One of the primary techniques for measuring metallicity evolution utilizes observations of rest-frame optical nebular emission lines from bright H II regions within high-redshift actively star-forming galaxies (SFG).  The relative strengths of these lines provide a diagnostic tool to identify both the abundance of heavy elements and the ionization parameter (U), which is a dimensionless value typically defined by the ratio of the volume of ionizing photon flux and hydrogen density. While powerful, this method is challenging for redshifts z $\gtrsim 1$, since the surface brightness of galaxies fades dramatically, and optical emission lines are redshifted into the near-infrared (NIR), where ground-based observations are hindered by higher atmospheric background. Only within the last decade, with the advent of more sensitive near-infrared spectrographs on 8-10m class telescopes, has there been an attempt at specifically studying high redshift (z $\gtrsim$ 1) galaxies using optical emission lines for metallicity diagnostics (e.g., \citealt{shap04, shap05, sav05, erb06a, maier06, kriek07, liu08, maio08,hain09}).  A significant number of these galaxies have nebular emission line ratios that fall out of the ``normal" distribution of local star forming galaxies and it has been speculated that either the physical conditions (ie, density, ionization parameter) of star formation at early epochs may differ from local SFGs \citep{liu08,brinch08,hain09} and/or these high-z object's spectra are composites of both active star formation and active galactic nuclei (AGN) emission \citep{groves06}.  

Metallicity studies typically target galaxies that have no signs of AGN activity, as determined by their SED (i.e., from HST and Spitzer imaging) and their rest-frame UV spectroscopy. However, there is growing evidence for AGN activity in some SFGs (or UV selected galaxies) from optical narrow and broad emission lines in z$\gtrsim$ 2 galaxies \citep{kriek07, hain09, shapiro09, fink09}.  In addition, evidence for weak AGN activity has recently been discovered in two z$\sim$1.6 star-forming galaxies observed with an integral field spectrograph coupled with an adaptive optics (AO) system \citep{wright09}. These galaxies have high [N II]/H$\alpha$ ratios concentrated into very small regions ($\lesssim$ 0$\farcs$1). Without the high spatial resolution achieved with adaptive optics and the 2D mapping capabilities of integral field spectrographs, these weak AGN would not have been discovered, since none of these galaxies have SEDs or UV spectra indicative of hosting an AGN.  In this paper, we present further evidence to support the presence of weak AGN in star-formation dominated galaxies with the detection of spatially  concentrated \oiii~emission for the galaxy HDF-BMZ1299 (z$\sim$1.5985).  We generate 2D nebular ratio maps for HDF-BMZ1299, and for the first time are able to use multiple aperture sizes for a high-z galaxy to place this galaxy on the traditional \citealt{bald81} (BPT) diagram.  An estimated AGN contribution is derived, and we show that integrated nebular ratios can be greatly affected by the observed PSF and aperture width (e.g., slit size).  Throughout the paper we use the concordance cosmology \citep{kom09}.

\section{Observations \& Data Reduction}\label{observ}

HDF-BMZ1299 (hereafter BMZ1299) was first observed at the W.M. Keck II 10m telescope using the near-infrared integral field spectrograph OSIRIS (OH Suppressing Infrared Imaging Spectrograph: \citealt{larkin06}) with the LGS AO system \citep{wiz06} on May 20, 2008.  We detected \ha~and \nii~emission in these observations, which are presented in \citet{wright09}.  Subsequent observations of \oiii~emission were obtained with OSIRIS on August 21, 2008 and are presented here for the first time.  The coarsest scale (0\farcs1) with the narrowband filter Jn3 (1.1 \micron) was used to measure \oiii~emission across a 4$\farcs$8 x 6$\farcs$4 field of view to encompass the entire galaxy and to allow dithering within OSIRIS's lenslet array.  Five individual 900 second exposures were taken of BMZ1229 to yield a total integration time of 1.25 hours.  An \citet{elias82} standard, HD106965, was observed for flux calibration.

\begin{figure*}[t]
\epsscale{1.0}
\begin{center}
\plotone{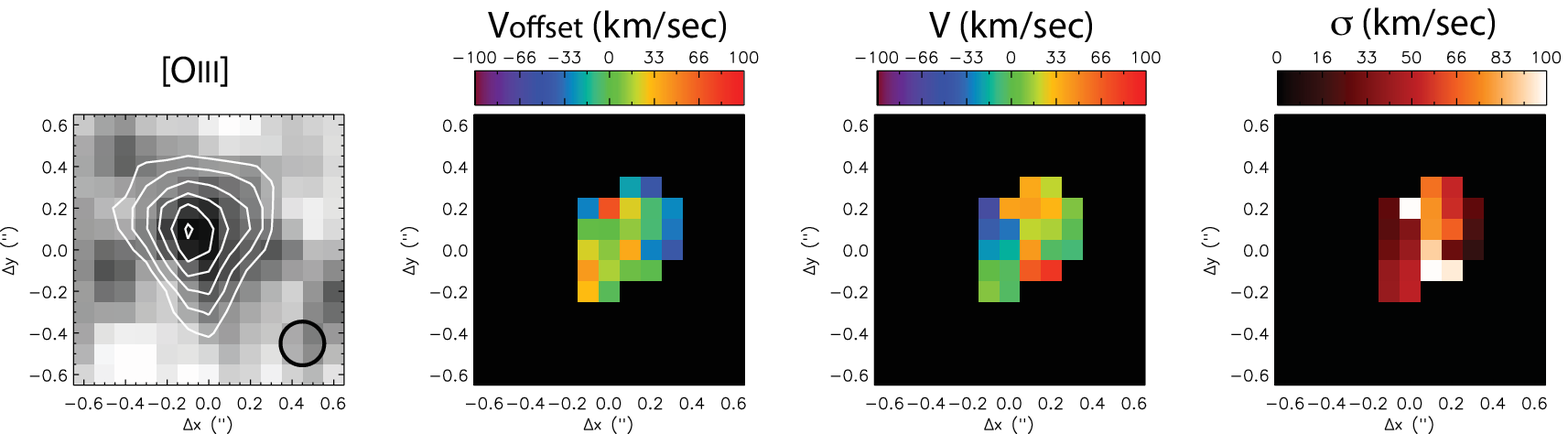}
\caption{\linespread{1}\small\textit{ 
Two-dimensional \oiii~flux distribution and kinematics of BMZ1299.(Left) Integrated \oiii~flux observed from OSIRIS with \ha~contours overlaid. The FWHM of the tip-tilt star is overlaid as a black circle demonstrating the AO performance. (Middle left) Two-dimensional \oiii~velocity offset relative to \ha~velocity profile. (Middle right) Two-dimensional \oiii~kinematics showing spatial distribution of velocity centers (km s$^{-1}$). (Right) Two-dimensional \oiii~kinematics showing the velocity dispersion (km s$^{-1}$) map. 
 }}\label{f1}
\end{center}
\end{figure*}

The OSIRIS data reduction was performed as described in \citet{wright09}, and in addition we used custom IDL routines for cleaning residual bad pixels in the reduced cube, scaling sky subtraction with atmospheric OH lines near the observed \oiii~emission, and flux calibrating.  A final reduced integrated \oiii~image of BMZ1299 is presented in Figure \ref{f1} with \oiii~radial and dispersion velocities that are further described in Section \ref{dyn}.  A wavelength solution was refined for both our H-band and J-band observations using OH sky lines to align the \ha~and \nii~reduced cube and the \oiii~reduced cube to similar velocity bins \footnotemark.  For kinematic and nebular ratio analysis, we spatially smoothed the \oiii~final mosaicked image in each wavelength channel with a Gaussian kernel (FWHM = 0\farcs2) to achieve higher signal-to-noise (S/N) and to match the resolution of the \ha~and \nii~mosaicked cube (see \citealt{wright09}).   Both cubes were spatially aligned by matching the location of the peak \oiii~emission to the \ha~peak emission (we discuss implications of this assumption in Section \ref{ratio}).  

\footnotetext{The deviation of the previous wavelength
  solution between H and J band was $\sim$1 spectral channel
  (or $\Delta$$\lambda$$\sim$0.15 nm in J-band), which was due
  primarily to differences in the temperature of the grating
  at the time of each observation.}

\section{Nebular Ratios}\label{ratio}

As described in Section \ref{observ}, the \oiii~reduced cube is aligned spatially and spectrally to the \ha~and \nii~reduced cube with matching velocity resolution.  Using a range of velocity bins, we generate nebular ratio maps for N2H$\alpha$ (N2H$\alpha$ = log (\nii/\ha)), O3H$\beta$ (O3H$\beta$ = log(\oiii/\hb)), and O3N2 (O3N2 = log(\oiii/\nii)).  We use photometric data and spectral population fitting analysis (see \citealt{erb06b}) to derive an E(B-V) extinction for BMZ1299 (E(B-V)=0.35), from which we estimate dereddened emission line fluxes and nebular ratios.  \hb~was not directly detected, and therefore we use the dereddened \ha~emission to infer the amount of \hb~emission for BMZ1299 assuming case B recombination (F$_{H\beta}$ = F$_{H\alpha}$/2.86). This is a conservative limit for the \hb~flux, since larger extinction would decrease the estimated \hb~flux and therefore increase the overall O3H$\beta$ ratio. In addition, if we assume \hb~does not arise from \hii~regions but instead is dominated by the hard-ionizing radiation from an AGN, then the estimated amount of \hb~flux would decrease as well (F$_{H\beta}$ = F$_{H\alpha}$/3.1 for an AGN, \citealt{ost89}), and again would increase the overall O3H$\beta$ values.  

\begin{figure*}[t]
\epsscale{1.0}
\begin{center}
\plotone{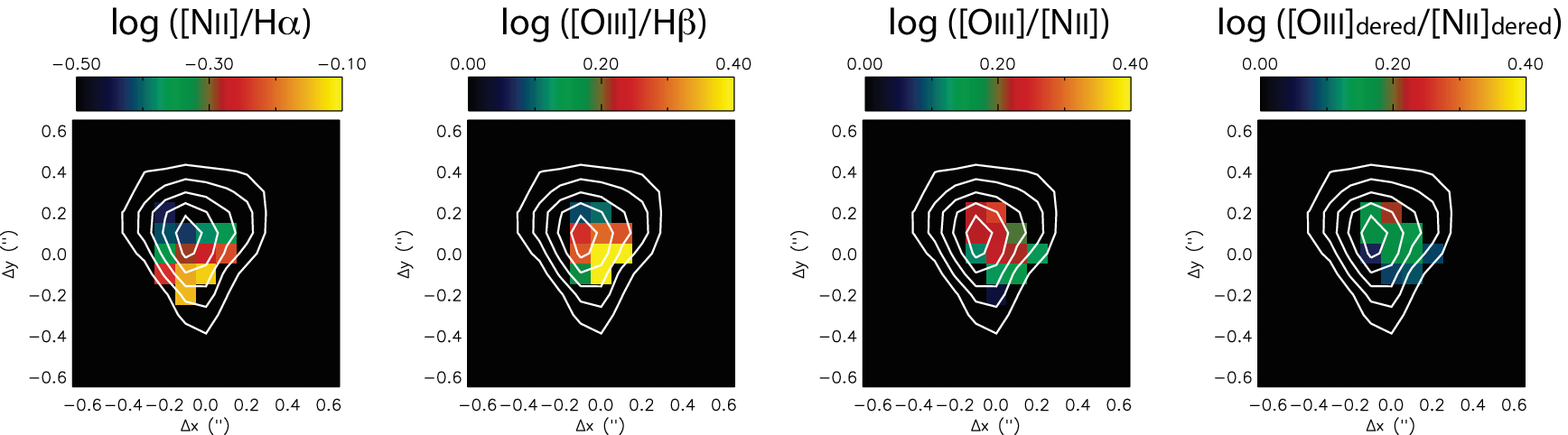}
\caption{\linespread{1}\small\textit{ 
\nii/\ha, \oiii/\hb, and \oiii/\nii~ratio maps for BMZ1299 are presented to illustrate locations of high and low nebular ratios. Contours of \ha~emission are superimposed on each image. All ratio maps are generated with a matching velocity bin ($\sim$70 \kms). Note that in \citet{wright09} \nii/\ha~ratio maps are plotted at differing velocity bins and widths to illustrate locations of peak ratios. Dereddened ratios are only plotted for \oiii/\nii since the dereddned values for both \nii/\ha~and \oiii/\hb~ratios are equivalent to the observed ratios. There are high \nii/\ha~and \oiii/\nii~ratios concentrated in one to two spatial elements in similar spatial locations in BMZ1299. These spatially concentrated ratios are plotted in the BPT diagram in Figure \ref{f3} and \ref{f4}, and are found to lie in the distribution of local AGN emission.
}}\label{f2}
\end{center}
\end{figure*}

Figure \ref{f2} shows the 2D nebular ratio images for BMZ1299 of N2H$\alpha$, O3H$\beta$, O3N2, and dereddened O3N2. Since a number of assumptions are made for dereddening the fluxes (e.g., global extinction across the galaxy), we are only presenting dereddened values for O3N2 where extinction has an impact on the observed ratios.    We use the average extinction errors associated with the BM/BX sample in \citealt{erb06a} to yield an associated extinction error for BMZ1299. This extinction error is used for all dereddened emission line fluxes and is included in error propagation.  As presented in Figure \ref{f2}, there are high ratios for N2H$\alpha$ and O3H$\beta$ that are spatially concentrated within one to two spatial elements (0\farcs2, 1-2 kpc) in the galaxy, that are of order FWHM of the PSF observed for both \ha~and \oiii~observations.  

For comparison with our values, we have selected $\sim$26,000 emission-line objects from the Sloan Digital Sky Survey (SDSS\footnotemark) fourth data release (DR4: \citealt{adel06}). We applied several criteria to ensure a sample of local high S/N emission line objects: (1) all SDSS objects are between of 0.005 $\lesssim$ z $\lesssim$ 0.25; (2) all objects have detected \hb, \oiii, \ha, \nii~emission that have S/N $\gtrsim$ 10; and (3) objects that were used in \citet{kauf03b}. These SDSS emission line objects are plotted on the traditional BPT diagram (O3H$\beta$ vs. N2H$\alpha$) in Figure \ref{f3} and \ref{f4}.  Emission-line diagnostic figures are used to empirically distinguish between emission generated from either star formation or AGN activity.  For additional comparison, previous results from high-z metallicity studies that have detected all four emission lines are also in Figure \ref{f3} and \ref{f4} \citep{shap05,maier06,erb06a,liu08,hain09}. Figure \ref{f3} illustrates nebular ratios for all the high redshift individual galaxies.  Figure \ref{f4} illustrates the average binned nebular ratios for each high redshift study.  The emission line ratios from \citet{liu08} represent several galaxies binned at two different redshifts (z=1.0 and z=1.4), which are derived from \citet{shap05} and \citet{liu08}. For both \citet{maier06} and \citet{erb06a}, we have taken average values of each emission line flux measurement and propagated their respective errors to derive a single binned measurement for all their observed galaxies. From \citet{hain09}, the ``Clone" galaxy (z=2.0) nebular ratio from their full aperture Keck NIRSPEC spectrum is presented in both Figure \ref{f3} and \ref{f4}.

We show three different ratios in Figure \ref{f3} and \ref{f4} for BMZ1299, each of which is calculated over a different spatial area of the galaxy.  For a conservative comparison, all \oiii/\hb~ratio values presented for BMZ1299 are 2$\sigma$ lower limits given the assumptions described previously for estimating the \hb~flux.  The nebular line ratio from BMZ1299's spatially concentrated region (0\farcs2x0\farcs2) lies within the local SDSS distribution of AGN (open circle on Figure \ref{f3} and \ref{f4}). As described in Section \ref{observ}, if our alignment method of peak-to-peak \oiii~and \ha~emission is not accurate, then the spatially concentrated O3H$\beta$ ratio would only increase, moving it higher into the AGN distribution.  Therefore, aligning the peak emission between these two emission lines serves as a lower limit for the O3H$\beta$ ratio.  Integrating over a seeing-sized region, however, like a slit spectrograph, we find that the line ratios of BMZ1299 lie within the local SDSS distribution of star forming galaxies (filled circle on Figure \ref{f3} and \ref{f4}).  The difference between concentrated and integrated ratios are 0.32 dex for N2H$\alpha$ and 0.34 dex for O3H$\beta$.  
We can conservatively remove the central source by scaling the PSF from the tip-tilt star to match the PSF of the central source, and apply an aperture correction to the AGN emission based on the fraction of flux in the same region of the tip-tilt star. The resulting line ratios for the extended source are consistent with local star forming galaxies (square+circle on Figure \ref{f3} and \ref{f4}).  

The evidence that the \oiii~emission is significantly more compact than \ha~emission is also striking since the AO performance (and hence Strehl ratio) in J-band, where \oiii~is observed, is poorer than H-band, where \ha~is observed. If the galaxy's PSF has a larger FWHM than the tip-tilt star, then the extended emission line ratios would move further towards the bulk of local SFGs (moving down and to the left on Fig \ref{f4}). Therefore, we find that the source is well described by an extended region of star formation consistent with local SFGs, and a central AGN. When observed as an integrated source, this combination of star formation and AGN activity produces line ratios that are intermediate between local star forming dominated and AGN  dominated galaxies, and matches other high redshift sources that have not been spatially resolved. 

Using the emission line ratios from the integrated emission from BMZ1299 with the central source removed, we are able to make a rough metallicity estimate of BMZ1299.  We use the two indicators from \citet{pett04}: 12 + log(O/H) = 8.90 + 0.57 $\times$ log(\nii/\ha)~and 12 + log(O/H) = 8.73 - 0.32 $\times$ log(\oiii/\nii), where 8.66 represents a solar oxygen abundance.  We find that both indicators yield consistent results, with 8.56$\pm$0.07 and 8.67$\pm$0.08 from \nii/\ha~and \nii/\oiii, respectively.  This implies that BMZ1299 has an upper limit of slightly subsolar to solar metallicity.  We caution that the errors on these metallicities could be underestimated, as they do not account for assumptions in using a global extinction and our PSF scaling estimates for the central source removal. The stellar population fitting of BMZ1299's SED yields a stellar mass of 1.3x10$^{10}$ \msun \citep{wright09}.  BMZ1299 shows evolution from the local MMZ relationship \citep{trem04}, but it is difficult to make comparisons to other evolved MMZ relationships at higher redshift (e.g., z$\sim$1-1.4 galaxies in \citealt{liu08}, z$\sim$2.2 galaxies in \citealt{erb06a}) since our metallicity estimates only serve as an upper limit.

\footnotetext{Retrieved from http://www.mpa-garching.mpg.de/SDSS/}

\section{\oiii~Dynamical Properties}\label{dyn}

At each spaxial where \oiii~has a S/N $\gtrsim$5, a Gaussian was fitted to the emission line to derive a 2D spatial map of radial velocities and velocity dispersions. Figure \ref{f1} contains 2D images of the integrated \oiii~flux distribution with \ha~contours overlaid, \oiii~velocity offset with respect to \ha~emission, \oiii~radial velocity profile, and \oiii~velocity dispersion.  The overall velocity profile of \oiii~is flat across the galaxy and resembles the observed velocity profile of the \nii~emission. In contrast, the \ha~emission is spatially extended and shows a 2D velocity gradient of $\pm$150 \kms~that is well-fit to an inclined disk model with low intrinsic velocity dispersion \citep{wright09}. \oiii~global velocity dispersion is 56$\pm$13  \kms~and is comparable to \nii~velocity dispersion of 71$\pm$10.  The spatial compactness, lack of velocity gradients, and similar velocity dispersions in both \oiii~and \nii~emission provide further support for the presence of an AGN. 

\section{Discussion}\label{discuss}

We have presented OSIRIS LGS-AO \oiii~observations that, combined with previous observations of \ha~and \nii~of BMZ1299, allow us to place the resolved substructure of a high-redshift galaxy on the traditional BPT diagram for the first time. These spatially resolved observations are further evidence to support the presence of AGN activity in an otherwise star-formation dominated high-reshift galaxy, as first presented in \citet{wright09}.  Our observations of BMZ1299 have provided the first direct evidence that its integrated spectrum is a composite with contributions from both \hii~regions and AGN activity. 

\begin{figure*}[t]
\epsscale{0.8}
\begin{center}
\plotone{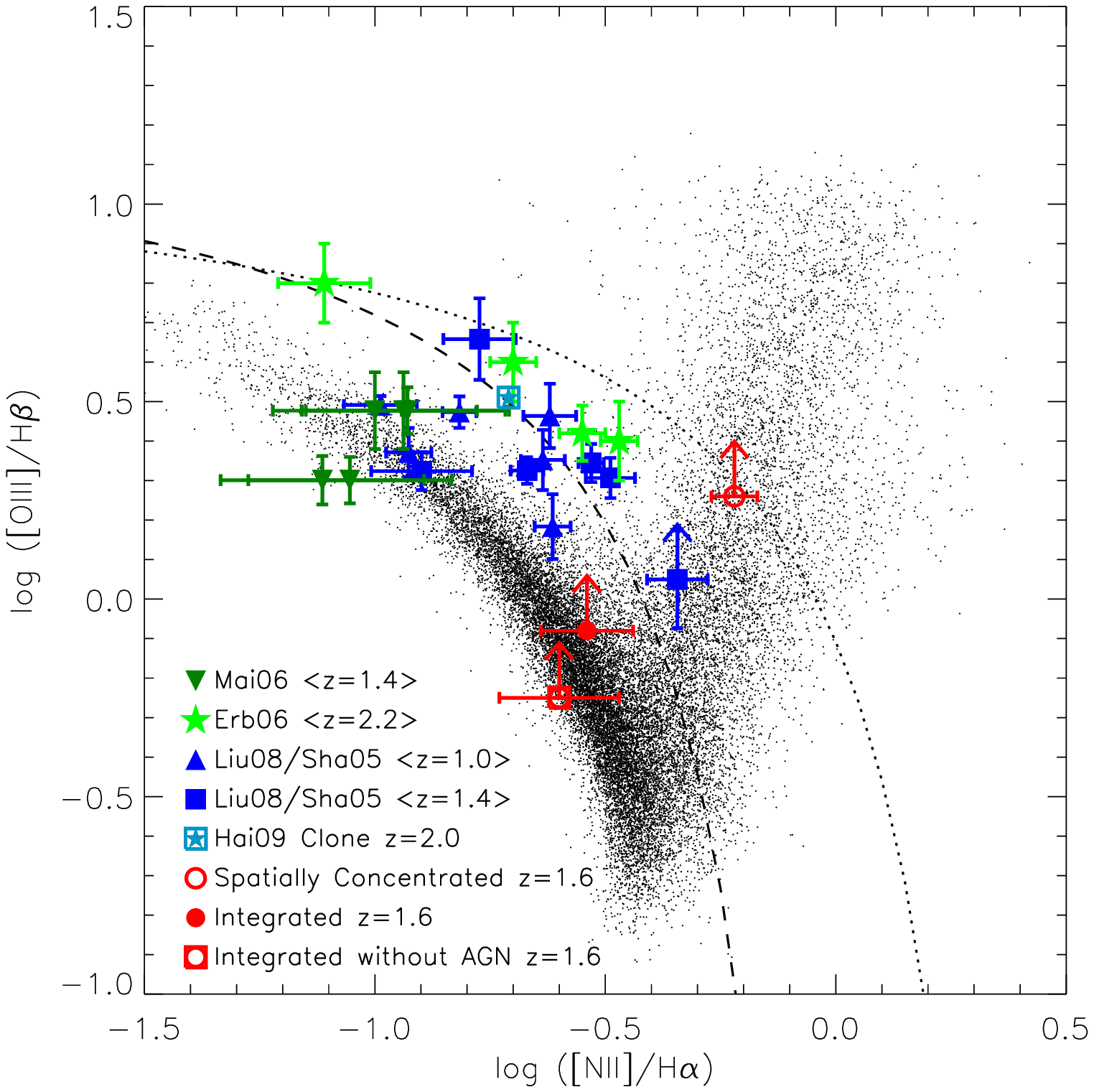}
\caption{\linespread{1}\small\textit{ 
H II region and AGN diagnostic diagram for the emission ratios of log([OIII]/\hb) vs. log([NII]/\ha). SDSS local galaxies and AGN (0.05 $\lesssim$ z $\lesssim$ 0.25) are represented with black points, which illustrate the tight star forming galaxy sequence in the bottom left of the figure and the AGN Seyfert and LINER sequence rising to the top right. The dashed line is the empirical curve from \citet{kauf03a} separating the local SDSS galaxies from AGN. The dotted line is the theoretical curve from \citet{kew01} representing the limit for star forming galaxies generating line emission from H II regions. Previous long-slit spectroscopy observations of individual high redshift galaxies with their emission line ratios; dark green upside down triangle for z$\sim$ 1.4 from \citet{maier06}, green star for z$\sim$2.2 from \citet{erb06a}, blue triangle for z$\sim$1.0 and blue square for z$\sim$1.4 from both \citet{shap05} and \citet{liu08}. Emission line ratios for BMZ1299 are over-plotted in red to illustrate how spatially concentrated and integrated ratios across the galaxy are highly dependent on the observed PSF.  All log(\oiii/\hb) values for HDF-BMZ1299 are plotted as 2$\sigma$ limits; increasing the assumed extinction of this source would increase the \oiii/\hb~ratios. The open red circle lying in the AGN SDSS distribution is the spatially concentrated ratios from a 0\farcs2 x 0\farcs2 region of BMZ1299 (as seen in Figure \ref{f2}). The solid red circle represents the integrated ratios from the entire spatial extent of BMZ1299. The red square with open circle are the ratios for the integrated galaxy with estimated contribution of the AGN emission removed. 
 }}\label{f3}
 \end{center}
 \end{figure*}

\begin{figure*}[t]
\epsscale{0.8}
\begin{center}
\plotone{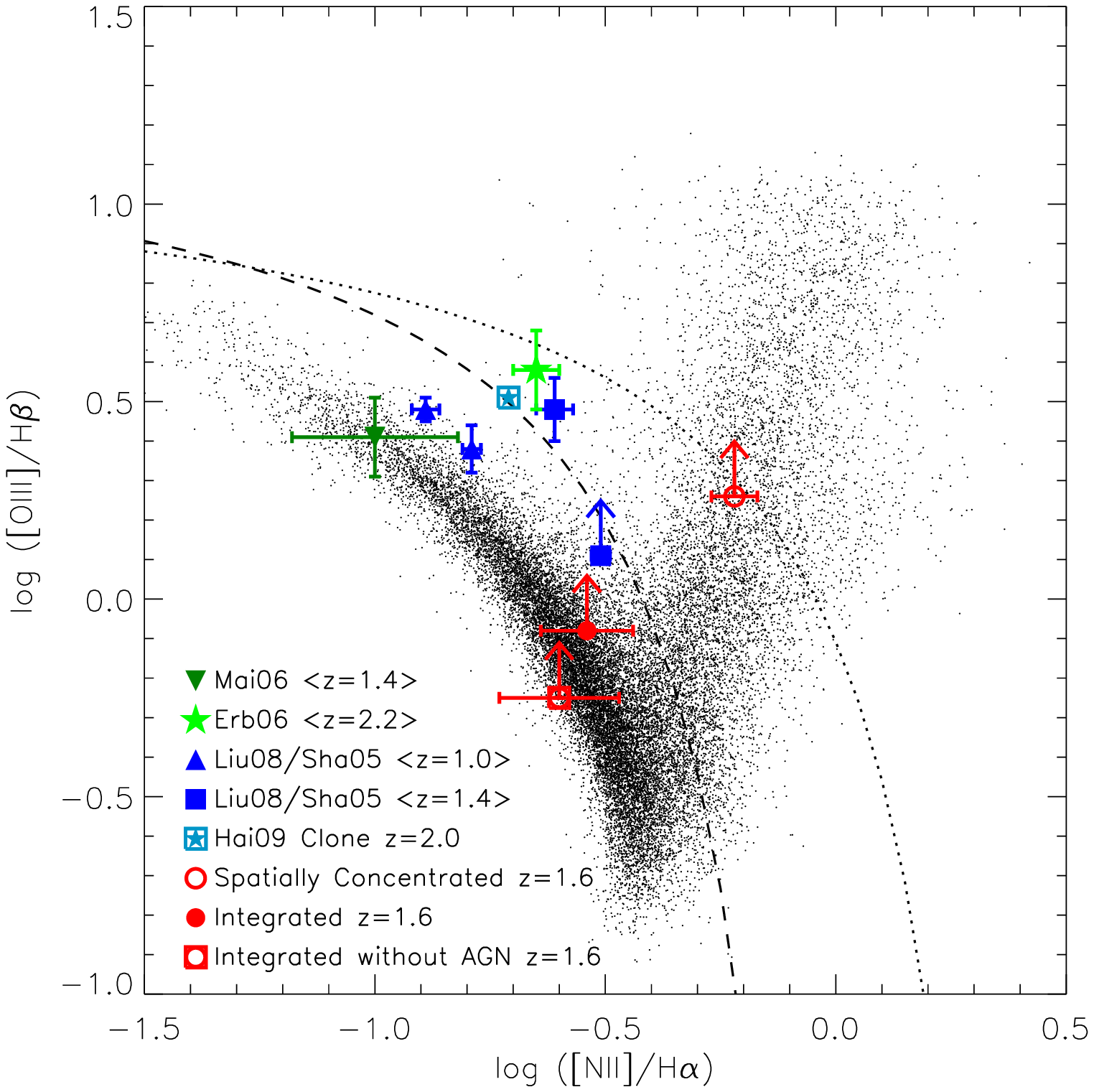}
\caption{\linespread{1}\small\textit{ 
Similar to Figure \ref{f3}.  Instead of individual high redshift galaxies, previous long-slit spectroscopy emission line ratios are binned and presented; dark green upside down triangle for z$\sim$ 1.4 from \citet{maier06}, green star for z$\sim$2.2 from \citet{erb06a}, blue triangle for z$\sim$1.0 and blue square for z$\sim$1.4 from both \citet{shap05} and \citet{liu08}. All log(\oiii/\hb) values for HDF-BMZ1299 are plotted as 2$\sigma$ limits; increasing the assumed extinction of this source would increase the \oiii/\hb~ratios. The open red circle lying in the AGN SDSS distribution is the spatially concentrated ratios from a 0\farcs2 x 0\farcs2 region of BMZ1299 (as seen in Figure \ref{f2}). The solid red circle represents the integrated ratios from the entire spatial extent of BMZ1299. The red square with open circle is the ratios for the integrated galaxy with estimated contribution of the AGN emission removed. 
 }}\label{f4}
 \end{center}
 \end{figure*}

Previous seeing-limited, long-slit spectrograph observations have found that nebular ratios of some high-z galaxies are offset from the local SFG sequence and lie within the ``transition" region of the BPT diagram.  Roughly 45\% of high-z galaxies on the BPT diagram lie above the empirical demarcation from \citealt{kauf03a} (dashed line in Figure \ref{f3} and \ref{f4}) that distinguishes between nebular ratios from star formation and AGN.  One proposed explanation for this offset (as seen in \citealt{shap05,erb06a,kriek07,liu08,hain09}), is that physical conditions of star formation at high-redshift are different from those in local galaxies.  The most notable differences between high-redshift SFGs and local SFGs are that high-z sources tend to have higher SFRs (10-100 \myr) and are morphologically compact \citep{law07}. \citealt{liu08} investigated the characteristics of local SDSS galaxies that were located in ``offset" regions above and below the theoretical \citet{kew01} star formation limit, and concluded that the majority of the offset sources would have differing \hii~physical conditions caused by a significantly larger ionization parameter from increased electron temperatures and densities. \citet{brinch08} explored this further by investigating the specific SFR (SFR/M$^{*}$) of local SDSS star forming galaxies, and found that local SFGs with higher specific SFR were ``offset" above the local SFG distribution.  They also suggested that these offsets are due to higher ionization parameters from increased electron densities or a non-negligible amount of escaping Lyman continuum photons.  \citet{brinch08} constrained their analysis to local SFGs that were located below the empirical demarcation from \citealt{kauf03a} and did not probe the more elevated ``transition" objects in SDSS that mimic where other high-z galaxies' integrated nebular ratios are located.

As an alternative explanation to differing physical conditions at high redshift, it was first suggested by \citet{groves06} that these observed ``offset" high-redhshift SFGs were perhaps a composite of nebular emission arising from both star formation and AGN activity that mimic local transition objects seen in SDSS.  The aperture sizes used in both SDSS and high-z spectroscopic observations are quite large, since typical SFGs are 1$\arcsec$ in size and long slit spectrographs observe the entire galaxy ($\sim$ 8.5 kpc at z=1.5), and 3$\arcsec$ fiber apertures in SDSS are used for spectroscopic measurements ($\sim$5 kpc at z=0.1).  For a composite object, the location of such a source on an N2H$\alpha$ and O3H$\beta$ BPT would be affected by multiple factors: (1) relative emission-line strengths from both star formation and AGN activity; (2) metallicities within both the star forming regions and the AGN; and (3) observed PSF and aperture size used for deriving the nebular ratios of the galaxy.  The location of the galaxies observed by \citet{erb06a,liu08,hain09} lie in a regime with significantly lower N2H$\alpha$ values compared to locally found composite objects, and it was speculated by \citet{groves06} that these galaxies are composite objects but of lower metallicity than local SDSS galaxies.  \citealt{liu08} investigated the  characteristics of local SDSS galaxies that were located in ``offset" regions above and below the theoretical \citealt{kew01} star formation limit, and estimated that 20\% of the anomalous emission in high-redshift galaxies could be due to AGN contamination.

We are able to estimate the contributing flux from the AGN by scaling the tip-tilt star PSF to that of BMZ1299's \ha~and \oiii~observations.  We find an estimated AGN luminosity of L$_{H\alpha}$ = 3.7$\pm$0.5x10$^{41}$ erg s$^{-1}$ and  L$_{[OIII]}$ = 5.8$\pm$1.9x10$^{41}$ erg s$^{-1}$, which is three orders of magnitude fainter than typical luminosities of observed high-redshift quasars \citep{mcint99} at this epoch.  These AGN luminosities are more comparable to the higher end of the local Seyfert population luminosities \citep{hao05}.  A local analog for these \ha~and \oiii~AGN luminosities are those observed from NGC 4151 (e.g.,\citealt{heck05}).  Recently, studies have shown a weak correlation between \oiii~and X-ray luminosities of local Type I and II AGN \citep{heck05,georg09}.  Using this empirical correlation, we infer a predicted range of X-ray luminosity L$_{X (2 - 10 keV)}$ $\sim$ 10$^{41 - 43}$ erg s$^{-1}$ for BMZ1299.  This range agrees well with the 3$\sigma$ X-ray detection limit of BMZ1299 within the Chandra and Hubble Deep Field North of L$_{X (2 - 8 keV)}$ $\lesssim$ 2.3x10$^{42}$ erg s$^{-1}$ \citep{alex03}.  Given observed high-redshift L$_{X (2 - 10 keV)}$ luminosities (e.g., \citealt{main02, silver08}), BMZ1299 would harbor one of the weakest AGN discovered at this epoch.  

The AGN luminosity function is predicted to peak at z$\sim$1-3 for a range of AGN luminosities \citep{franca05,rich06} .  These studies have shown that brighter luminosity AGN peak at higher redshift and have lower number densities compared to fainter luminosity AGN which peak at lower redshifts.  For an L$_{X (2 - 10 keV)}$ $\sim$ 10$^{42-43}$ erg s$^{-1}$ AGN at z=1.5, the comoving number density is predicted to flatten around $\sim$5-8x10$^{-4}$ h$^{-3}$ Mpc$^{-3}$  \citep{franca05}.  If we make an exploratory assumption that AGN with similar luminosities reside in halos typical of BMZ1299, then given the number density of BM objects (5$\pm$2.5x10$^{-3}$ h$^{-3}$ Mpc$^{-3}$ :\citealt{adel05}), roughly 5-35\% of star forming galaxies at z$\sim$1.5 would host an AGN of this luminosity. This estimate serves as a lower limit since it does not account for lower and higher luminosity AGN that could be present within the same galaxy population. This illustrates that the number density of the faint-end luminosity AGN at z $\gtrsim$ 1 could account for a significant fraction of high-redshift star forming galaxies that are found in the transition region of the BPT diagram.  

Although we have only presented one source, these spatially resolved observations have shown the presence of composite spectra at high-redshift, and that AGN would offer an explanation for the higher ionization parameters and electron densities observed. Since our AO observations are biased to detect bright compact emission, we may also be discovering only the sources that are composite high-redshift spectra (as predicted from \citealt{liu08}). Both interpretations of differing \hii~physical conditions and composite (\hii--AGN) spectra may prove necessary to explain the entire high redshift galaxy population.  Regardless, AO coupled with an IFS has proven to be invaluable for disentangling composite (\hii~- AGN) galaxy spectra, and may play an essential role in future discoveries of lower luminosity high-z AGN and their potential feedback on galactic formation.

\acknowledgements
Data presented herein were obtained at W.M. Keck Observatory, which was made possible by generous financial support from the W.M. Keck Foundation. The authors would like to acknowledge the dedicated members of the Keck Observatory staff, particularly Jim Lyke, Randy Campbell, and Al Conrad,  who helped with the success of our observations. We would also like to acknowledge Elizabeth Barton for helpful conversations and support.  Research was supported by funding from a NASA HF-51265.01 fellowship. The authors wish to recognize the significant cultural role and reverence that the summit of Mauna Kea has always had within the indigenous Hawaiian community. We are most fortunate to have the opportunity to conduct observations from this ``heiau" mountain.


\begin{deluxetable}{lc}
\tabletypesize{\scriptsize}
\tablecaption{HDF-BMZ1299}
\tablewidth{0pt}
\tablehead{
\colhead{Properties} & 
\colhead{Value}  \\
}
\startdata 
z$_{[OIII]}$ & 1.5985 \\
F$_{[OIII]}$ \tablenotemark{a} & 11$\pm$1x10$^{-17}$ \\
F$_{H\alpha}$ \tablenotemark{b} & 22$\pm$1x10$^{-17}$ \\
F$_{[NII]}$ \tablenotemark{b} & 6.4$\pm$1.5x10$^{-17}$ \\
\cutinhead{Integrated line ratios}
log(\nii/\ha)$_{inte}$ & -0.54$\pm$0.10 \\
log(\oiii/\hb)$_{inte}$ & $\gtrsim$ -0.08\tablenotemark{c} \\
log(\oiii/\nii)$_{inte}$ & 0.24$\pm$0.11 \\
log(\oiii$_{dered}$/\nii$_{dered}$)$_{inte}$ & 0.18$\pm$0.19 \\
\cutinhead{Spatially concentrated ratios for 0\farcs2 x 0\farcs2 region \tablenotemark{d}}
log(\nii/\ha)$_{conc}$ & -0.22$\pm$0.05 \\
log(\oiii/\hb)$_{conc}$ & $\gtrsim$ 0.26\tablenotemark{c} \\
log(\oiii/\nii)$_{conc}$ & 0.10$\pm$0.04  \\
log(\oiii$_{dered}$/\nii$_{dered}$)$_{conc}$ &0.05$\pm$0.04 \\
\cutinhead{Integrated ratios with AGN emission removed}
log(\nii/\ha)$_{no AGN}$ & -0.60$\pm$0.13 \\
log(\oiii/\hb)$_{no AGN}$ & $\gtrsim$ -0.25\tablenotemark{c} \\
log(\oiii/\nii)$_{no AGN}$ & 0.18$\pm$0.15 \\
log(\oiii$_{dered}$/\nii$_{dered}$)$_{no AGN}$ & 0.16$\pm$0.25 \\
\cutinhead{Estimated AGN emission \tablenotemark{e}}
F$_{H\alpha AGN}$ & 2.2$\pm0.3$x10$^{-17}$  \\
L$_{H\alpha AGN}$ & 3.7$\pm$0.5x10$^{41}$  \\
F$_{[OIII] AGN}$  & 3.4$\pm$1.1x10$^{-17}$  \\
L$_{[OIII] AGN}$ & 5.8$\pm$1.9x10$^{41}$\\
\enddata
\tablenotetext{a}{Integrated Fluxes (ergs s$^{-1}$ cm$^{-2}$) observed in Jband.}
\tablenotetext{b}{Integrated Fluxes (ergs s$^{-1}$ cm$^{-2}$) observed in Hband \citep{wright09}.}
\tablenotetext{c}{2$\sigma$ limit for  \oiii/\hb~nebular ratio}
\tablenotetext{d}{Ratios from $\Delta$$\lambda$$\sim$0.4nm.}
\tablenotetext{e}{Estimated fluxes (ergs s$^{-1}$ cm$^{-2}$) and luminosity (ergs s$^{-1}$) for AGN.}
\label{table}
\end{deluxetable}

\clearpage
\end{document}